\documentclass[1p,sort&compress]{elsarticle}

\usepackage{hyperref}
\usepackage[utf8]{inputenc}
\usepackage{graphicx}
\usepackage{color}
\usepackage{amsfonts,amsthm,amsmath,amssymb}
\usepackage{bm,bbm}

\newcommand{\be}{\begin{equation}}
\newcommand{\ee}{\end{equation}}
\newcommand{\ba}{\begin{eqnarray}}
\newcommand{\ea}{\end{eqnarray}}
\def\bs{\begin{subequations}}
\def\es{\end{subequations}}
\renewcommand{\leq}{\leqslant}
\renewcommand{\geq}{\geqslant}
\def\a{\alpha}
\def\b{\beta}
\def\de{\delta}
\def\g{\gamma}
\def\la{\lambda}

\def\ve{\varepsilon}
\def\Om{\Omega}
\def\om{\omega}
\def\De{\Delta}

\def\s{\sigma}

\def\cX{\mathcal{X}}

\def\bD{\mathbbm{D}}
\def\ds{d_{\rm S}}
\def\dh{d_{\rm H}}

\def\p{\partial}

\newcommand{\Eq}[1]{(\ref{#1})}
\def\com{\color{magenta}}
\def\cob{\color{blue}}
\newcommand{\book}[5]{{#1}, #2, #3, #4, #5}
\newcommand{\books}[4]{{#1}, #2, #3, #4}
\newcommand{\oarX}[1]{\href{http://arxiv.org/abs/#1}{{\ttfamily\com arXiv:#1}}}
\newcommand{\arX}[1]{\href{http://arxiv.org/abs/#1}{{\ttfamily\com arXiv:#1}}}
\newcommand{\doin}[6]{\href{http://dx.doi.org/#1}{{\cob #2 #3 {#4} (#6) #5}}}
\newcommand{\doinn}[5]{\href{http://dx.doi.org/#1}{{\cob #2 {#3} (#5) #4}}}
\newcommand{\doij}[5]{\href{http://dx.doi.org/#1}{{\cob #2 #3 (#5) #4}}}

\newcommand{\ndoinn}[5]{\href{#1}{{\cob #2 {#3} (#5) #4}}}

\newcommand{\tia}[1]{#1,}

\def\lp{\ell_{\rm Pl}}
\def\tp{t_{\rm Pl}}
\def\mpl{m_{\rm Pl}}

\def\rme{e}
\def\rmd{d}
\def\rmi{i}
\def\bd{\mathbbm{d}}
\def\bD{\mathbbm{D}}

\begin{document}

\begin{frontmatter}

\title{Dimensional flow and fuzziness in quantum gravity: emergence of stochastic spacetime}

\author{Gianluca Calcagni}
\ead{calcagni@iem.cfmac.csic.es}
\address{Instituto de Estructura de la Materia, CSIC, Serrano 121, 28006 Madrid, Spain}

\author{Michele Ronco}
\ead{michele.ronco@roma1.infn.it}
\address{Dipartimento di Fisica, Universit\`a di Roma ``La Sapienza'', P.le A.\ Moro 2, 00185 Roma, Italy}
\address{INFN, Sez.\ Roma1, P.le A.\ Moro 2, 00185 Roma, Italy}

\begin{abstract}
We show that the uncertainty in distance and time measurements found by the heuristic combination of quantum mechanics and general relativity is reproduced in a purely classical and flat multi-fractal spacetime whose geometry changes with the probed scale (dimensional flow) and has non-zero imaginary dimension, corresponding to a discrete scale invariance at short distances. Thus, dimensional flow can manifest itself as an intrinsic measurement uncertainty and, conversely, measurement-uncertainty estimates are generally valid because they rely on this universal property of quantum geometries. These general results affect multi-fractional theories, a recent proposal related to quantum gravity, in two ways: they can fix two parameters previously left free (in particular, the value of the spacetime dimension at short scales) and point towards a reinterpretation of the ultraviolet structure of geometry as a stochastic foam or fuzziness. This is also confirmed by a correspondence we establish between Nottale scale relativity and the stochastic geometry of multi-fractional models.
\end{abstract}




\end{frontmatter}


\section{Introduction}

After many years of research, we are not yet close to an acknowledged unique quantum theory of gravity, partly because of the lack of experimental guidance. The mathematical and conceptual challenges raised by the attempt of combining quantum-mechanical and general-relativistic principles produced plenty of different approaches to the problem of quantum gravity (QG) \cite{Ori09,Fousp,Smo17}. Among them, we count string theory \cite{Zwi09}, the tripod of group field theory, loop quantum gravity and spin foams \cite{rov07,thi01,Per13,GiSi}, causal dynamical triangulation \cite{AGJL4}, causal sets \cite{Dow13}, asymptotically safe gravity \cite{LaR5,NiR,RSnax}, non-commutative spacetimes \cite{ADKLW,BIMM} and non-local quantum gravity \cite{Tom97,Mod1,BGKM,CaMo2}, just to mention some of the most popular models available in the literature. Over the last twenty years, this considerable theoretical effort has started both to figure out phenomenological predictions that could be tested with the presently-achievable levels of experimental sensitivity and to gradually focus on few results that seem independent of the specific quantum-gravity framework adopted \cite{gacLRR}. In fact, even if there are great differences between inequivalent approaches, some common features have been noticed. One of the most recurrent findings in the field is \textit{dimensional flow} (or dimensional running), i.e., a change of spacetime dimension with the scale of the observer. In almost all quantum-gravity models, the dimensionality of spacetime exhibits a dependence on the scale, changing (or ``flowing'') from the topological dimension $D$ in the infrared (IR) to a different value in the ultraviolet (UV). There can be more than a single relevant scale and, thus, the dimension can change many times before reaching its far-UV value at a scale that is often identified with (or recognized as) the Planck length $\lp = (G\hbar/c^3)^{1/(D-2)}$. Sometimes, the concept of dimension does not even survive deep into these UV scales and it dissolves into some highly non-smooth structure (for instance, multi-fractal, discrete, or combinatorial). All known quantum gravities are \emph{multi-scale} by definition because they all have an anomalous scaling of the dimension \cite{tH93,Car09,fra1} (see \cite{NewRev,Car17} for a scan of the literature and more and newer references). A recent strategy for easily realizing the running of the dimension has been followed by \textit{multi-fractional theories}, comprehensively reviewed in \cite{NewRev}. In these models, the basic ingredient implementing dimensional flow is a non-trivial factorizable integration measure
\begin{equation*}
\rmd q^0(x^0) \rmd q^1(x^1)\cdots \rmd q^{D-1}(x^{D-1}) = \frac{\rmd q^0}{\rmd x^0} \rmd x^0 \frac{\rmd q^1}{\rmd x^1} \rmd x^1\cdots\frac{\rmd q^{D-1}}{\rmd x^{D-1}} \rmd x^{D-1}\,.
\end{equation*}
The profiles $q^\mu(x^\mu)$ are determined uniquely and solely by requiring to reach the IR limit as an asymptote \cite{first}. An approximation of the full measure, which will be of interest here, is the so-called binomial space-isotropic profile
\begin{equation}
\label{multimeas}
q^\mu(x^\mu) \simeq (x^\mu-\overline{x}^\mu) + \frac{\ell_*}{\a_\mu}\left|\frac{x^\mu-\overline{x}^\mu}{\ell_*} \right|^{\a_\mu}\,,\qquad \a_\mu=\a_0,\a.
\end{equation}
Here there is no summation over the index $\mu = 0, \dots, D-1$. The fractional exponents $0 < \a_0,\a < 1 $ are directly related to both the spectral and the Hausdorff dimensions ($\ds$, $\dh$) at very short distances $\ell \lesssim \ell_*$; if $\a_0=\a$, then $\ds\simeq D\a\simeq\dh$ in the UV for the theories considered here (with fractional or $q$-derivatives \cite{NewRev}). In the above measure, we are assuming spatial isotropy (same $\alpha$ for all space directions) and the existence of only one characteristic length $\ell_*$. These approximations can be relaxed without difficulty but, since the full exact form of the measure $q^\mu(x^\mu)$ is not needed here,\footnote{The binomial measure is the approximation of a measure with many scales smaller than $\ell_*$, all of which are effectively ``screened'' by $\ell_*$ and that do not appear in the phenomenology for all practical purposes \cite{first}.} for the sake of our argument we will limit our attention to \Eq{multimeas}, at least at the beginning. The binomial measure \Eq{multimeas} is obtained by a coarse-graining procedure from the most general case of measures with logarithmic oscillations, that contain at least another shorter length $\ell_\infty \leq \ell_*$ and a frequency $\omega$ \cite{frc2}. Later on, we will consider also 
this case. 

Both $\alpha$ and $\ell_*$ are free parameters of the theory with the only constraints that $\ell_*$ is expected to be small in order to respect experimental constraints on the dimension of spacetime (typically, $\ell_*$ is much smaller than the electroweak scale \cite{NewRev}) and $\alpha$ must stay in the interval $\alpha \in (0,1)$ for arguments of theoretical consistency \cite{NewRev}. The second flow-equation theorem \cite{first} or rigorous arguments of multi-fractal geometry \cite{frc2} fix the measure $q^\mu(x^\mu)$ uniquely, but not the physical frame where measurements are performed. In fact, while a multi-fractional geometry is designed to adapt with the scale of the observation, our devices (rods, clocks, and so on) are not.\footnote{The contrary happens in asymptotic safety, where measuring devices are adaptive (momenta acquire a scale dependence from the renormalization group flow) \cite{fra7}.} This is realized at the price of breaking Poincaré invariance, so that physical observables have to be computed in a fixed preferred frame. This poses the so-called problem of the choice of {presentation}, which consists in the choice of $\bar{x}^\mu$. Although there are infinite possible choices, four are special \cite{trtls} and the second flow-equation theorem reduces them to two \cite{first}.
 
In this paper, we show that the limitations on the measurability of spacetime distances, which have been obtained by many authors combining quantum mechanical (QM) and general relativistic (GR) arguments \cite{padma,gara,alu,ngdam,amelino} or relying on specific quantum-gravity models \cite{Hooft,ven,kon,ellis,loop}, can be regarded as a multi-scale effect. In fact, multi-fractional theories naturally carry an additional non-trivial contribution to the magnitude of a distance, which is not present in a classical theory with standard integration measure. For special values of $\alpha$ in Eq.\ \eqref{multimeas}, this multi-fractional contribution can be reinterpreted as an intrinsic uncertainty (or \emph{fuzziness}, in QG jargon) on the measurement of spacetime distances exactly of the same type encountered in a standard (i.e., non-multi-scale) model where both QM and GR are taken into account \cite{ngdam,amelino}. This suggests that classical multi-fractional models in Minkowski spacetime (i.e., in the absence of curvature) partially encode both QM and GR effects, and that they do so thanks to dimensional flow. This correspondence between semi-classical quantum gravity and multi-fractional theories will allow us to give a physical interpretation to the ambiguities of the multi-fractional theories with fractional and $q$-derivatives. In fact, the comparison of the multi-fractional uncertainty on the distance with two different lower bounds found by Ng and Van Dam \cite{ngdam} and Amelino-Camelia \cite{amelino} will select two preferred values for the fractional exponent, $\alpha = 1/3$ or $\alpha = 1/2$. Remarkably, the second value was recognized as special since early papers \cite{fra1,frc1,frc2} for several theoretical reasons \cite{NewRev}, including its frequency of appearance in the quantum-gravity landscape of theories. Moreover, we will identify $\ell_*$ with the Planck length $\lp$ in the former case, while in the latter we will obtain $\ell_* = \lp^2 / s < \lp$, where $s$ is the observation scale. Interestingly, in the second case the dependence on the scales at which the measurement is being performed becomes explicit. This is exactly what is expected to happen in multi-fractal geometry and, in particular, in multi-fractional theories, where the results of measurements depend on the observation scale. In our analysis, this effect comes directly from equating the multi-fractional uncertainty with the semi-classical one. Turning this sort of duality around, we solve the long-standing presentation problem in a surprising way. Consider a length $L$ in a multi-fractional spacetime with binomial measure (the same argument holds for time intervals). The typical difference between $L$ and the value $\ell$ that would be measured in an ordinary space is \cite{trtls}
\be\label{ll0}
|L-\ell|\simeq \de L_\a:=\frac{\ell_*}{\a}\left(\frac{\ell}{\ell_*}\right)^\a\,.
\ee
Until now, the multi-fractal correction $\de L_\a$ has been regarded as a deterministic effect signaling an anomalous scaling at scales $\lesssim \ell_*$. Here, we reinterpret it as an intrinsic uncertainty of measurements, so that lengths cannot be measured with a precision smaller than $\de L_\a$. This reinterpretation is not arbitrary and will rely on the so-called harmonic structure of geometry, associated with a deep UV discrete scale invariance generating an infinite hierarchy of scales. This structure is at the core of a precise relation between Nottale scale relativity \cite{Not93,Not97,Not08} and a multi-fractional measure that is nowhere differentiable, a property which is distinctive of stochastic geometries.

After reviewing the distance and time uncertainty estimates of \cite{ngdam,amelino} in section \ref{revi}, we will obtain them directly in multi-fractional stochastic spacetimes in section \ref{mai}. Section \ref{disc} is devoted to a discussion of the consequences of the main results for multi-fractional theories and quantum gravity at large, also comparing with previous attempts to relate dimensional flow and fuzziness. A condensed presentation can be found in \cite{ACCR}.


\section{Distance and time uncertainty estimates}\label{revi}


\subsection{Review of the estimates}

We begin by reviewing the Salecker--Wigner procedure \cite{SW} for the quantum measurement of spacetime distances and highlight how, taking into account the quantum nature of measuring devices, the presence of gravitational interactions forbids to identify a length with arbitrarily good accuracy (zero uncertainty). A necessary observation is that QM and GR give completely different definitions of the position of an object. In the former, it is simply identified by its four coordinates $x^\mu$, but there is no prescription for the actual measurement of these coordinates. On the contrary, coordinates have no meaning by themselves in GR and, in order to identify a ``position'' (a spacetime event), one has to specify an operational procedure to measure the distance between the observer and the measured object. Thus, for the purpose of measuring a given distance, Salecker and Wigner \cite{SW} recognized three basic devices: a clock, a light signal, and a mirror. We set the initial time when the light ray leaves the clock site. Then, it is reflected by the mirror at a distance $L$. When the light ray comes back to the clock, the time we read is  $T = 2L/c$, where $c$ is the speed of light. Now, quantum mechanics affects this measurement by introducing an uncertainty $\delta L$. In the same way, if we try to measure the time of travel $T$, the latter will be affected by a quantum uncertainty $\delta T$. To calculate these uncertainties, we follow two possible lines of reasoning. The first, due to Ng and Van Dam \cite{ngdam}, seeks the major element of disturbance for the measurement of both distance and time in the QM motion of the quantum clock. The second argument, by Amelino-Camelia \cite{amelino}, focuses on the QM uncertainty in the position of the center of mass of the whole system. In both cases, since we are considering QM properties of devices, the system is initially described by a wave packet with uncertainties on position and velocity that affect the measurement by producing an initial spread $\delta L(0)$. Then, the length $L$ acquires an uncertainty $\delta L(T) \simeq \delta L(0) + \delta v(0) T$, where $\delta v(0)$ is the QM uncertainty on the velocity of the system (there is a slight difference in the two cases, since in the first one $\delta v (0)$ refers to the clock, while in the second to the center of mass), over the duration $T$ of our measurement. We discuss explicitly the uncertainty on length measurements but an analogous argument applies also to time measurements, for which there is an equivalent result. First, let us follow the approach of Ref.~\cite{ngdam}. As aforementioned, the uncertainty $\delta L$ is induced by the fact that, as a quantum object, the clock cannot stay absolutely still. It has a QM uncertainty  on its velocity $\delta v(0) = \delta p(0)/M \geq \hbar /[2 M \delta L(0)]$, where $M = M_{\rm c}$ is the mass of the quantum clock. In the light of this, we can rewrite the QM uncertainty on the measurement of our distance as
\begin{equation}
\delta L(T) \geq \delta L(0) + \frac{\hbar L}{cM\delta L(0)} \geq  \frac{\hbar L}{cM\delta L(T)}\,,
\end{equation} 
where we have replaced $T = 2L/c$ and also maximized the denominator by putting $\delta L(T)$ in place of $\delta L(0)$. (Due to the quantum motion of the clock, the uncertainty on the length measurement is expected to increase, i.e., $\delta L(T) \geq \delta L(0)$.) Therefore, using only standard QM arguments, we find
\begin{equation}
\label{qmunc}
(\delta L)^2 \geq  \frac{\hbar L}{cM}\,.
\end{equation}
Now we add GR effects. Turning gravity on, we know that the gravitational field of the clock will affect the measurement of the distance $L$. As soon as gravity comes into play, spacetime is no longer Minkowski and, thus, distances change due to curvature effects.  How much does this modify the distance we are measuring? To answer that, one can calculate the uncertainty $\delta L$ produced by the gravitational field of the clock. Suppose our quantum clock is spherically symmetric and that the metric around it is approximately Schwarzschild. Passing to ``tortoise coordinates'' \cite{fink}, the time interval for a complete trip is given by
\begin{equation*}
\widetilde{T} = T + \frac{r_{\rm S}}{c}\left[\ln\left |\frac{r_c+L}{r_{\rm S}}-1\right | -\ln\left |\frac{r_c}{r_{\rm S}}-1\right |\right] \,,
\end{equation*}
 where $r_{\rm c}$ is the size of the clock and $r_{\rm S} = 2GM_{\rm c}/c^2$ is the Schwarzschild radius. Then, the distance reads
\begin{equation*}
\widetilde{L} = L + \frac{r_{\rm S}}{2}\ln\left |\frac{r_c+L-r_{\rm S}}{r_c-r_{\rm S}}\right| \,.
\end{equation*} 
Here, the first term is the distance in Minkowski spacetime, while the second contribution is the gravitational correction due to the clock. Thus, we have
\begin{equation*}
\delta L \simeq \frac{r_{\rm S}}{2}\ln\left |\frac{r_c+L}{r_c}\right|
\end{equation*}
in the approximation $r_c \gg r_{\rm S}$. This expression tells us that, having introduced GR effects, there is an additional uncertainty to the measurement of the distance given by 
\begin{equation*}
\delta L \geq \frac{GM_{\rm c}}{c^2},
\end{equation*}
having neglected the numerical factor $\ln[(r_c+L)/r_c]$.
Combining this bound with the QM one of Eq. \eqref{qmunc}, we finally obtain \cite{ngdam}
\begin{equation}
\label{NgDa}
\delta L \geq \delta L_\frac{1}{3} := (\lp^2 L)^\frac{1}{3} \,,
\end{equation}
where the subscript stresses that this lower bound has exponent $1/3$. Following a similar reasoning, one can easily find an intrinsic uncertainty also on measurements of time intervals \cite{ngdam}:
\begin{equation}
\label{NgDaT}
\delta T \geq \delta T_\frac{1}{3} :=  (\tp^2 T)^\frac{1}{3} \,.
\end{equation}

The argument by Amelino-Camelia \cite{amelino} is slightly different. In that case, one identifies the source of disturbance with the center of mass of the system rather than with the clock. The QM part of the  reasoning remains the same, the only difference being the replacement of $M_{\rm c}$ with the total mass $M_{\rm tot}$ into Eq.\ \eqref{qmunc}. On the gravity side, we simply require that the total mass is not large enough to form a black hole, i.e.,  $M_{\rm tot}\leq c^2 s / G$, where $s$ is the size of the total system made up of the clock plus the light signal plus the mirror. In fact, if a black hole formed, then the light signal could not propagate to the observer, thereby making the measurement impossible. Combining this restriction with the QM uncertainty, one finds \cite{amelino}
\begin{equation}
\label{AmCam}
\delta L \geq  \delta L_\frac{1}{2} := \sqrt{\frac{\lp^2 L}{s}} \,;
\end{equation}
 the subscript $1/2$ is to distinguish the exponent of the uncertainty. Analogously, the uncertainty on time measurements reads \cite{amelino}
\begin{equation}
\label{AmCamT}\
\delta T \geq \delta T_\frac{1}{2} :=   \sqrt{\frac{\tp^2 T}{t}} \,,
\end{equation}
where $t = s/c$.


\subsection{\texorpdfstring{QM$+$GR$=$QG}{}: the physics of quantum gravity emerges}

There are at least two comments to make concerning expressions \eqref{NgDa} and \eqref{AmCam}. Similar considerations apply also to Eqs.\ \eqref{NgDaT} and \eqref{AmCamT}. First,  they both depend on the time $T = 2L/c$ of the measurement, a feature that has been often regarded as a sign of quantum gravitational decoherence. Second and most importantly for what follows, it is worth noting that the interplay of QM and GR principles determines a feature that, hopefully, might help our intuition on the physics of QG. In fact, one ends up with an intrinsically irreducible uncertainty on the measurement of a single observable, in this case the distance or time interval. The combination of QM and GR affects geometric observables, such as distance and time, and this was often interpreted as a confirmation that QG requires a new understanding of geometry (as explicit constructions of quantum gravity eventually confirmed). This single-observable uncertainty is not just a QM effect, since QM only imposes a limitation on the simultaneous measurement of conjugate variables. It also has no counterpart in GR. In fact, one recovers the standard case $\delta L = 0$ by turning off either GR or QM.  As far as we consider only QM limitations, we can of course get $\delta L = 0$ by taking the infinite-mass limit $M_{\rm c},M_{\rm tot}\rightarrow \infty$ in Eq. \eqref{qmunc}. However, this is no longer possible when we consider GR interactions since, in the presence of gravity, the apparatus would form a black hole before reaching an infinite mass. Again, from Eq. \eqref{qmunc} one can see that the uncertainty on the distance $L$ goes to zero if we turn off QM by sending $\hbar \rightarrow 0$. Moreover, both $\delta L_\frac{1}{3}$ and $\delta L_\frac{1}{2}$ depend on $\lp$, that goes to zero if one takes either the limit $G \rightarrow 0$ (i.e., we neglect gravity) or $\hbar \rightarrow 0$ (i.e., we neglect quantum properties). However, as soon as both QM and GR effects are taken into account, there is an irreducible $\delta L$. These uncertainty expressions are telling us that quantum gravity might require either a new measurement theory or an exotic picture of spacetime, or both. In the second case, we are led to expect a sort of spacetime foam at scales close to the Plack distance. In fact, the appearance of a limitation on the measurement of distances suggests that, at Planckian scales, spacetime is no longer the smooth continuum we are used to in both QM and GR. At those very-high-energy (very-short-distance) scales, the presence of an intrinsic $\delta L$  may mean that spacetime is made of events that cannot be localized with arbitrary sharpness. In QG, classical continuous spacetime is replaced by a fuzzy structure.

All these considerations, born of the heuristic arguments of \cite{ngdam,amelino}, later found confirmation in concrete QG theories \cite{Ori09,Fousp}, each of which realizes this irregular UV structure in very different ways \cite{tH93,Car09,NewRev}. One may ask whether and why, if real QG can be embodied by so many diverse theories, the heuristic combination of QM and GR leading to Eqs.\ \eqref{NgDa}--\eqref{AmCamT} is essentially correct. In the next section, we answer to this question as follows: the heuristic arguments are correct and part of the reason is that they rely, inadvertently, on a universal feature of quantum gravities: dimensional flow.

To show this, we shall analyze the measurement of a distance in the multi-fractional theories with fractional or $q$-derivatives in flat embedding space adding neither QM nor GR effects. Despite these notable absentees, we will see that a non-trivial contribution to distance measurements is present and it has the same structure of the spatial uncertainties \eqref{NgDa} and \eqref{AmCam} of the semi-classical QG arguments we just reviewed. The same is true also for the time uncertainties \eqref{NgDaT} and \eqref{AmCamT}. Interpreting the multi-fractional correction to spacetime distances as an uncertainty, not only are we able to fix the ambiguities of the model (i.e., its free parameters $\alpha$ and $\ell_*$ as well as the presentation), but we also recognize how multi-fractional models intrinsically unite the combination of GR and QM effects. This provides further support to the view that multi-fractional theories can be regarded both as stand-alone proposals and as effective models of QG, and that they are able to capture at least two of (what we think to be) the characteristic features of quantum geometry, namely, dimensional flow and spacetime fuzziness. The latter concept, typically ambiguous when not applied to a particular theory \cite{MoN}, will be given a precise meaning later.\footnote{In theories with discrete pre-geometric structures, such as the set GFT-LQG-spin foams, ``fuzziness'' means combinatorial and discreteness effects \cite{COT2,COT3}.}


\section{Fuzziness, dimensional flow and stochastic geometry}\label{mai}

Let us now show that the bounds \eqref{NgDa}--\eqref{AmCamT} on the measurement of spacetime distances can be reinterpreted as purely classical multi-fractional effects in the absence of gravity. Classical multi-fractional theories encode both QM and GR effects, which we have used above to obtain the uncertainties $\delta L$ and $\delta T$ in a semi-classical setting with elementary notions of QM and GR on a standard geometry with local measure $dx^0dx^1$ (the $D$-dimensional case is straightforward).\footnote{In multi-fractional theories, one replicates the same argument for all directions separately and combines everything into the distance $\ell=\sqrt{\ell_1^2+\ell_2^2+\ldots}$. Alternatively, one can pick a multi-scale measure dependent only on the Lorentz distance; these models are purely phenomenological in general, but they capture the correct dimensional flow \cite{fra1,fra2,fra3}.} On one side of the correspondence, we have a multi-fractional theory with two structures: a built-in dimensional flow, which is a feature usually derived (rather than assumed) in top-down approaches to QG, and a stochastic-spacetime structure we will describe in this section. On the other side, there is an uncertainty on distance measurements, a property that follows from a naive bottom-up approach combining just QM and GR principles without adding any hypothetical QG ingredient. The correspondence states that this uncertainty is reproduced by dimensional flow plus an intrinsic randomness at the microscopic level. As surprising as it may be, both dimensional flow and spacetime fuzziness can be obtained at the same time as a result of having a deformed non-trivial integration measure. This may explain the common origin of these two QG features despite strong differences among different QG approaches. Apparently, when gravity is quantized one always obtains a multi-scale geometry and some sort of ``irregular'' or non-smooth structure in the UV. Conversely, a multi-scale geometry and a UV fuzzy structure naturally lead, when properly defined, to the unification of quantum mechanics with gravity.

The interpretation we propose here has also the advantage of drastically reducing the ambiguities of multi-fractional models. In fact, by comparing the multi-fractional uncertainty to the bounds of Eqs.\ \eqref{NgDa}--\eqref{AmCamT} we succeed in fixing the free parameters $\alpha$ and $\ell_*$. In particular, the multi-fractional length $\ell_*$ turns out to be related to the Planck length $\lp$, a fact that strengthens the interpretation of multi-fractional theories as QG descriptions. Moreover, the problem of having different presentation choices, discussed below, is either reinterpreted as an effect of underlying spacetime fuzziness or is irrelevant in the presence of such a structure.


\subsection{Deterministic view}

To this aim, we first comment briefly on the so-called presentation problem, which is part of the definition of a multi-fractional theory. We refer to \cite{NewRev} for a detailed discussion. The basic point is the following. In multi-fractional theories, the geometric coordinates $q^\mu(x^\mu)$ change with the scale (via their $\ell_*$ dependence), while fractional coordinates $x^\mu$ are scale-independent. The properties of experimental devices, used to take measurements, are independent of the observation scale. Thus, while the measure changes with the scale, clocks, rods and detectors do not. Consequently, physical observables have to be compared in the fractional frame with $x^\mu$ coordinates that do not adapt to the scale. This poses the problem of choosing a preferred fractional frame $\{x^\mu\}$ where Eq.\ \Eq{multimeas} is defined and observables are calculated. To say it in other words, geometric coordinates $q^\mu$ transform under the so-called $q$-Poincaré transformations ${q}^\mu({x'}^\mu)=\Lambda_\nu^{\ \mu}q^\nu(x^\nu)+a^\mu$ that are symmetries of the measure. However, physical quantities are determined in the fractional frame (which does not adapt to the scale, being it spanned by the $x^\mu$), where the dynamics is not invariant under these transformations. For this reason, in order to define physical observables it is necessary to fix a frame. It turns out that the frame ambiguity can be encoded in the vector parameter $\bar x^\mu$ in \Eq{multimeas}. Different presentation choices produce different measurement outcomes, corresponding to different theories with the same dimensional flow. We will show here that the presentation problem is not a problem at all when recognized as the source of an intrinsic uncertainty in the measurement of fractional distances. According to this perspective, the presentation ambiguity has the physical interpretation of an intrinsic spacetime distance fuzziness.

To this end, let us discuss the computation of a spatial distance. The difference between the spatial distance $\Delta q\geq 0$ expressed in terms of geometric coordinates $q^i$ ($i=1,\dots,D-1$) and the one $\Delta x\geq 0$ in fractional coordinates $x^i$ is encoded in the quantity $\cX := (\Delta q-\Delta x)/\Delta x$. Thus, the distance in the integer frame $\Delta q$ can be either larger or smaller than the distance $\Delta x$ measured in the fractional frame, depending on the sign of $\cX \lessgtr 0$. To explicitly see the influence of the presentation on the distance, let us consider a fractional frame labeled by $\bar{x}^\mu$. For simplicity but without loss of generality, let us consider one spatial dimension. It is not difficult to see that\footnote{This equation corresponds to \Eq{ll0}, where we improperly called $L=\Delta q$ a length (it is a geometric length in geometric coordinates).}
\begin{equation}\label{multdist}
\Delta q = \ell |1+\cX|\,,
\end{equation}
where $\ell = \De x=|x_{\rm B} - x_{\rm A} |$ and
\begin{equation}\label{chi}
\cX = \frac{1}{\alpha}\frac{\ell_*}{\ell} \left(\left | \frac{x_{\rm B}-\bar{x}}{\ell_*}  \right | ^\alpha -\left | \frac{x_{\rm A}-\bar{x}}{\ell_*}  \right | ^\alpha\right).
\end{equation}
From the above expressions, it is evident that different values of $\bar{x}$ (i.e., different presentations) give different results for the distance, but they do not change the anomalous scaling $\cX\sim x^\a$ solely governed by $\a$. Up to now, this was regarded as a freedom of the model and one had to make a choice of the presentation in order to have unambiguous predictions (\emph{deterministic view}).

Four special presentation choices have been identified as special among the others \cite{trtls}, but the second flow-equation theorem \cite{first} selects only two of these: the {initial-point presentation}, where $\bar{x} = x_{\rm A}$ (the presentation label is the beginning in time or space of the measurement, the zero of the clock or rod), and the {final-point presentation}, where $\bar{x} = x_{\rm B}$ (the end in time or space of the measurement, the number marked by the clock or rod when the experiment is over). In two of the three existing multi-fractional theories (the so-called theory $T_v$ with weighted derivatives and the theory $T_q$ with $q$-derivatives), it is not actually possible to give such a physical interpretation to the presentation choice as an intrinsic uncertainty, since none of these settings is invariant under translations and one cannot change $\bar x$ (a constant characteristic of the theory) at each experiment. Consider, for instance, the scalar-field action in $T_q$:
\be\label{Sphiq}
S_\phi=-\int\rmd^Dq(x)\,\left[\frac12\p_{q^\mu}\phi\,\p^{q^\mu}\phi+V(\phi)\right],\qquad \p_{q^\mu}:=\frac{\rmd}{\rmd q^\mu}=\frac{1}{(\p q)^\mu}\p_\mu\,.
\ee
The dynamics is not invariant under a shift $x^\mu\to x^\mu-\bar x^\mu$. On the other hand, in the third multi-fractional model, the theory with fractional derivatives (which we call $T_\g$ following \cite{NewRev}), the ordinary differential $\rmd$ is replaced everywhere by the differential $\bd$, an exterior multi-fractional derivative such that $\bd q^\mu(x^\mu)=q^\mu(\rmd x^\mu)$. The analogue of the scalar-field action \Eq{Sphiq} is
\be\label{Sphi}
S_\phi=-\int\bd^Dq(x)\,\left[\frac12\bD_\mu\phi\bD^\mu\phi+V(\phi)\right],\qquad \bD_\mu:=\frac{\bd}{\bd q^\mu}\,,
\ee
where we introduced the multi-scale derivatives $\bD_\mu$ proposed in \cite{NewRev} and $\bd^Dq(x)=\bd q^0(x^0)\cdots\bd q^{D-1}(x^{D-1})$ is the $D$-dimensional measure. At any plateau in dimensional flow (i.e., those scales where the spacetime dimension is approximately constant; in the binomial case \Eq{multimeas}, there are only two plateaux at $\dh\simeq D$ and $\dh\simeq D\a$), $\bd\simeq\bd q^\mu\p_\mu^\g\sim (\rmd x^\mu)^\a\p_\mu^\g$ coincides, when $\g=\a$, with the exterior derivative introduced in \cite{frc1} for a no-scale fractional measure, and the Euclidean distance is $\De(x,y)\simeq(\sum_\mu|y^\mu-x^\mu|^{2\a})^{1/(2\a)}$ in the deterministic view. It is not difficult to see that a shift $x^\mu\to x^\mu-\bar x^\mu$ leaves the derivatives $\bD_\mu$, the action \Eq{Sphi} and the equations of motion invariant.


\subsection{Stochastic view}

In \cite{trtls}, an analogy was noticed between the existence of different presentation choices in multi-fractional theories and the existence of different choices of the evaluation time of noise in stochastic processes. Consider a one-dimensional stochastic process $X(t)$ given by a noise with no deterministic component. In general, the graph $(t,X(t))$ is nowhere differentiable with probability 1 and one cannot write a meaningful differential $\rmd X(t)$ without some hand-made prescription on its inverse operation, integration \cite{Oks03,Gar04}. For an initial condition $X(t_{\rm i})$, we can make the splitting $\De t=t-t_{\rm i}$ as $t_{\rm i}=t_0<t_1<\cdots< t_{n-1}=t$ and write
\be\label{int0}
\int_{t_{\rm i}}^t\rmd X(t')\,f(t')=\lim_{n\to\infty}\sum_{j=0}^{n-1} f(\tilde t_j)\,[X(t_{j+1})-X(t_j)]\,,
\ee
for any test function $f$ in some suitably defined functional space. While the specific choice of the point $\tilde t_j\in [t_j,t_{j+1}]$ is irrelevant in the case of the Riemann--Stieltjes integral of an ordinary differentiable function $X(t)=x(t)$, it affects the output in the case of a process $X(t)$ fluctuating stochastically in $[t_j,t_{j+1}]$. The so-called It\^o and Stratonovich interpretations fix $\tilde t_j$ in two inequivalent ways (respectively, $\tilde t_j=t_j$ and $\tilde t_j=(t_{j+1}+t_j)/2$)
describing systems with different random properties. At the level of the Fokker--Planck equation \cite{KTH,Zwa01}, the It\^o--Stratonovich dilemma amounts to a choice of operator ordering in the Laplacian. In this case, the guiding principle is phenomenology: the stochastic system under examination will be better described by one choice instead of the other. In a multi-fractional particle-mechanics setting, the presentation problem precisely consists in the choice of $\tilde t_j\to\tilde t_j+\bar t_j$, where $X(t)$ is replaced by $q(t)$ \cite{trtls}. 

Having established that the presentation problem is basically equivalent to the It\^o--versus--Stratonovich prescription, we have two options. One is the \emph{deterministic view}: different choices correspond to different theories and only observations will be able to decide which prescription is correct. Perhaps, this view is not particularly elegant because it relies on an \emph{Ansatz} whose ultimate validity can be decided only by future experiments (how far in the future, we cannot tell). However, it is not particularly scandalous either, since it is not new in quantum gravity. Exactly the same It\^o--Stratonovich ambiguity in the Fokker--Planck equation appears in quantum cosmology, in the Fokker--Planck equation of eternal inflation \cite{LLM1} and in the Laplacian term of the Wheeler--DeWitt equation of canonical quantization \cite{De671,HPa}. In these cases, the guiding principle to fix the operator ordering is theoretical and can be more or less (but, more often than not, less) compelling.

In opposition to the deterministic view, the other option is more innovative. Generalizing to spacetime geometries, a nowhere-differentiable geometry can be realized in two ways: 
\begin{itemize}
\item by keeping the multi-fractional measure $q(x)$ deterministic but changing the differential calculus, or
\item by considering a nowhere-differentiable measure $q(x)$.
\end{itemize}
The first case corresponds to the theory $T_\g$, while the second case can be applied to all multi-fractional theories.

\subsubsection{Stochastic view with multi-fractional derivatives}\label{sec31}

Concerning the first possibility, fractional calculus embodies efficiently the nowhere differentiability typical both of multi-fractals \cite{KoG,RYS,RLWQ,NLM} and of anomalous stochastic processes or diffusion pseudo-processes \cite{MeK04,MeK05,Zas3,Sok12} (whose application to quantum gravity and multi-fractional theories can be found in \cite{frc4,frc7,CES}). The theory $T_\g$ relies on this calculus \cite{frc1} generalized to multi-fractional configurations such as \Eq{Sphi} \cite{NewRev}, where the whole integro-differential structure is deformed in such a way as to encode the irregularity property on a continuum. 

In the UV, $T_\g$ describes an irregular geometry very different from a smooth spacetime. Such irregularity is most naturally described in terms of probabilistic rather than deterministic features. For instance, we cannot exactly know what the dimension of spacetime is at a given scale, but we can find the most probable dimensions with a certain probability. This suggests to interpret the presentation ambiguity as a sign of a non-trivial stochastic-spacetime structure at microscopic scales, which cannot be classified or measured deterministically. Then, the initial-point and the final-point presentations (selected by the second flow-equation theorem) give us the two extreme values of the fluctuation interval of the fundamental uncertainty we would find in any measurement. This \emph{stochastic view} departs from the physical interpretation so far adopted in the literature, but for a good reason: it matches completely the heuristic arguments of the previous section which, in turn, permit to fix some of the free parameters of the multi-fractional measure.

In the theories $T_v$ and $T_q$ with a differentiable measure $q(x)$, one cannot realize \Eq{int0} straightforwardly because differential calculus is ordinary and one does not integrate over all possible labels $\bar t$ \cite{trtls}. However, it is possible to show that the multi-fractional derivatives $\bD_\mu$ can be approximated by the $q$-derivatives $\p/\p q^\mu(x^\mu)$ and that the propagators of $T_\g$ and $T_q$ agree in the UV \cite{NewRev}. Therefore, regarding $T_q$ as an approximation of $T_{\g=\a}$ carrying all the main features of the exact theory (same anomalous scaling in the UV, same scale hierarchy and value of $\ell_*$, and so on), one can investigate the effects of choosing the initial- or final-point presentation in the much simpler $T_q$, having always in mind that this is done only for technical simplicity. In the case of distance and time intervals such as those considered here, there is no difference between the two theories.

Taking the initial-point presentation, from Eqs.\ \eqref{multdist} and \eqref{chi} we get
\begin{equation}\label{delp}
\Delta q  = \left| \ell + \delta L_\a\right|\,,
\end{equation}
while, according to the final-point presentation, we obtain
\begin{equation}\label{delm}
\Delta q  = \left| \ell - \delta L_\a\right|\,,
\end{equation}
where we have defined
\begin{equation}
\label{multun}
\delta L_\a :=\ell\cX= \frac{\ell_*}{\alpha}\left(\frac{\ell}{\ell_*}\right)^\alpha.
\end{equation}
The initial-point presentation corresponds to a positive fluctuation $+\delta L_\alpha$, while in the final-point case one gets a negative fluctuation equal to $-\delta L_\alpha$. In the usual deterministic view, we would have one theory $T_\g^+$ or $T_q^+$ predicting $\De q> \ell$ (Eq.\ \Eq{delp}) physically inequivalent to another theory $T_\g^-$ or $T_q^-$ predicting $\De q< \ell$ (Eq.\ \Eq{delm}). In contrast, in the stochastic view the coexistence of the two allowed presentations is related to a limitation on the measurability of distances, and we do not have to decide a single presentation \emph{a priori}. In this way, an epistemological weakness of the model is overcome by replacing the idea of a UV geometry constituted by an anomalous spacetime where measurement can have arbitrary precision to a one where the UV limit is a fuzzy, or, better said, stochastic spacetime. The same discussion applies also to the time direction for which, in multi-fractional theories with fractional or $q$-derivatives, we have
\begin{equation}
\label{multiT}
\Delta q_0 = |T \pm \delta T_{\alpha_0}|\,, 
\end{equation}
with 
\begin{equation}
\label{deltaT}
\delta T_{\alpha_0} :=   \frac{t_*}{\alpha_0}\left(\frac{T}{t_*}\right)^{\alpha_0} \,,
\end{equation}
where $t_*$ is the time scale that characterizes the UV scaling of the time direction and $\a_0$ is the fractional exponent in the time direction $x^0$. 

Interpreting the presentation ambiguity as an intrinsic uncertainty in the determination of distances, Eq.\ \eqref{multun} tells us that a classical theory with a non-trivial measure, which exhibits a multi-scale (mono-scale, in the binomial case) behavior, naturally sets an obstruction on sharp measurements of distances as in a foam-like picture. A classical multi-scale theory in flat spacetime can reproduce the combined effect of GR and QM principles that, if held together, prohibit arbitrarily sharp measurements of spacetime intervals as we reviewed in section \ref{revi}. Besides resolving the presentation ambiguity, the interpretation of Eq. \eqref{multun} as a distance uncertainty also allows us to compare $\delta L_\alpha$ with the bounds $\delta L_\frac{1}{3}$ \eqref{NgDa} and $\delta L_\frac{1}{2}$ \eqref{AmCam} obtained by the naive combination of simple QM and GR arguments. Importantly, by comparing Eq.\ \eqref{multun} with Eq.\ \eqref{NgDa}, we can fix the multi-scale parameters $\alpha$ and $\ell_*$:
\begin{equation}\label{l13}
\delta L_\alpha = \delta L_\frac{1}{3}\quad \Longrightarrow\quad\alpha = \frac{1}{3}, \quad \ell_* = \lp\,.
\end{equation}
Thus, we get a preferred value for $\alpha$ and, what is more, the length scale $\ell_*$ of the model plays the role of the Planck length. This relation of $\ell_*$ with $\lp$ can be regarded as a confirmation that multi-fractional theories encode QG features in a highly non-trivial way. Similarly, identifying the multi-fractional fluctuation with the previously obtained semi-classical QG uncertainty, we also discover that the binomial measure should be isotropic in space and time, so that
\be
\a_0=\a\,. 
\ee
In fact, comparing Eqs.\ \eqref{deltaT} and \eqref{NgDaT}, we fix $\alpha_0$ and $t_*$ by 
\begin{equation}\label{t13}
\delta T_{\alpha_0} = \delta T_{\frac{1}{3}} \quad \Longrightarrow \quad \alpha_0 = \frac{1}{3}, \quad t_* = \tp \,,
\end{equation}
while confronting Eq.\ \eqref{multun} with \eqref{AmCam} and Eq.\ \eqref{deltaT} with \eqref{AmCamT} we get, respectively,
\ba
&&\delta L_\alpha = \delta L_\frac{1}{2}\quad \Longrightarrow\quad \alpha = \frac{1}{2}, \quad \ell_* = \frac{\lp^2}{s}\,,\label{l12}\\
&&\delta T_{\alpha_0} = \delta T_{\frac{1}{2}} \quad \Longrightarrow \quad \alpha_0 = \frac{1}{2}, \quad t_* = \frac{\tp^2}{t}\,.\label{t12}
\ea

To summarize, the theory with fractional derivatives describes spacetimes with a microscopic stochastic structure \cite{trtls}. The presentation label $\bar x^\mu$ prescribes how integrals on stochastic spacetime variables $X^\mu$ can be performed, and the presentation problem is similar to (not to say the same as) the It\^o--Stratonovich dilemma in stochastic processes. Inspired by this, instead of defining as many physically inequivalent theories (but with the same anomalous scaling) as the number of presentations and choosing one presentation among the others, we take all presentations at the same time. The measures $\{q^\mu(x^\mu)\,:\,\bar x^\mu \in\mathbbm{R}^D\}$ do not correspond to a class of (in)finitely many theories $T_\g^{\bar x}$ (labeled by $\bar x^\mu$) all with the same anomalous scaling: they are \emph{one} measure corresponding to \emph{one} theory $T_\g$ with an intrinsic microscopic uncertainty, limited by the initial- and final-point presentations. The theory $T_q$ with $q$-derivatives is related to $T_{\g=\a}$ by an approximation $\bd\simeq\rmd$ of the exterior derivative \cite{NewRev} and can be used to explore the physics of $T_{\g=\a}$. The multi-fractional theory $T_v$ is not an approximation of $T_\g$ nor has any stochastic microstructure, and we cannot juxtapose such a structure arbitrarily.

\subsubsection{Stochastic view with random measure}\label{sec32}

If we could make $q(x)$ nowhere differentiable, then we would be able to bypass the above limitations and extend the stochastic structure to all multi-fractional theories, not only to the one with fractional derivatives. To understand where the ``stochasticity'' could come from in classical multi-fractional spacetimes, it is useful to make a short digression and recall that the connection between a fractal and a stochastic structure in multi-scale spacetimes is not new. A proposal very similar to multi-fractional theories is Nottale's scale relativity \cite{Not93,Not97,Not08}, where lengths
\be\label{notL}
L=\ell+\zeta\ell_*\left(\frac{\ell}{\ell_*}\right)^\a
\ee
on a fractal spacetime are made of a deterministic differentiable part $\ell$ (the length on usual space) and a stochastic nowhere differentiable part. Here $\ell_*=1/\ve$ is the inverse of the resolution at which one is probing the geometry and $\zeta$ is a wildly fluctuating stochastic variable such that
\be\label{notlr}
\langle\zeta\rangle=0\,,\qquad \langle\zeta^2\rangle=\mp 1\,,
\ee
depending on whether the distance is time- or space-like. Because both scale relativity and multi-fractional spacetimes rely on a fractal geometry, these scenarios give about the same length expression. However, the original fractal-spacetime formulation of multi-fractional theories \cite{frc2} has been made much more solid thanks to a fundamental principle (slow IR dimensional flow) \cite{first} that reproduces the measure dictated by fractal geometry and, as we will see now, fixes some of the free parameters of scale relativity. In particular, not only is the stochastic random variable $\zeta$ of Nottale's ``fractal'' length $L$ present in a more general multi-fractional length if we go beyond the approximation \Eq{multimeas} of a binomial measure, but it is also fixed by the second flow-equation theorem, in contrast with the \emph{ad hoc} variable $\zeta$ in scale relativity. In fact, considering the second-order truncation of the full measure determined by the flow-equation theorem \cite{first}, we have (index $\mu$ omitted everywhere)
\bs\label{logos}
\begin{equation}\label{qF}
q(x)=x+\frac{\ell_*}{\a}\left|\frac{x}{\ell_*}\right|^\a F_\om(x)\,,
\end{equation}
where $F_\om(x)=F_\om(\la_\om x)$ is a complex modulation factor encoding a fundamentally discrete spacetime symmetry $x\to \la_\om x$ in the far UV ($\la_\om$ is fixed). Requiring the measure to be real-valued, one has \cite{first,NewRev,frc2}
\ba
F_\om(x) &=&\sum_{n=0}^{+\infty}F_n(x)\,,\\
F_n(x)&:=&A_n\cos\left(n\om\ln\left|\frac{x}{\ell_\infty}\right|\right)+B_n\sin\left(n\om\ln\left|\frac{x}{\ell_\infty}\right|\right),\label{logosb}
\ea\es
where $A_n$ and $B_n$ are constant amplitudes and $\lp\sim\ell_\infty\gtrsim\ell_*$. Since we will need some details about the derivation of this expression, let us make a short detour (in one dimension, for simplicity). The most general fractional complex measure $q(x)=x+\sum_n f_n(x)$ giving rise to a Hausdorff dimension slowly varying in the IR \cite{first} is given by the sum over $n$ of terms of the form 
\be\label{fen}
f_n(x)=\xi_n\left|\frac{x}{\ell_*}\right|^{\a_n+\rmi\om_n}+\eta_n\left|\frac{x}{\ell_*}\right|^{\a_n-\rmi\om_n},
\ee
where $\om_n>0$ and $\xi_n,\eta_n$ are constant. Assume, without loss of generality, that $\om_n=n\om$ (as in fractal and critical systems, as well as in quantum gravity as suggested by an analysis of the spacetime dimension \cite{cmplx}). Split $|x/\ell_*|^{\a_n\pm\rmi n\om}=c_{\pm n}|x/\ell_*|^{\a_n}|x/\ell_\infty|^{\pm\rmi n\om}$, where $\ell_\infty$ is an arbitrary length and
\be\label{ceqi}
c_{\pm n}=\left(\frac{\ell_\infty}{\ell_*}\right)^{\pm\rmi n\om}
\ee
is a pure phase. Then, $f_n(x)=|x/\ell_*|^{\a_n}(c_n \xi_n|x/\ell_\infty|^{\rmi n\om}+c_{-n}\eta_n|x/\ell_\infty|^{-\rmi n\om})$. We can reparametrize the system as
\be\label{cb}
c_n \xi_n=\frac{A_n-\rmi B_n}{2}\,,\qquad A_n\in\mathbbm{R}\,,\qquad B_n\in\mathbbm{R}\,.
\ee
If $\eta_n=\xi_n^*$ (real-valued $f_n$), then $f_n(x)=F_n(x)$ reproduces Eq.\ \Eq{logosb}.

The coordinate scaling ratio $\la_\om=\exp(-2\pi/\om)$ of the discrete scale invariance is governed by the frequency $\om$, which can be interpreted as the imaginary part of the Hausdorff dimension of spacetime \cite{cmplx}. The log-oscillating structure is typical of iterative (also called deterministic) fractals \cite{NLM,BGM1,DDSI,BGM2,DILu,BFSTV,ErEc,LvF}, complex and critical systems \cite{Sor98}, while in the context of quantum gravity it is solely determined by the flow-equation theorem \cite{first}. In multi-fractional theories, the modulation factor \Eq{logos} is usually approximated by only two frequencies, the zero mode $n=0$ ($F_0(x)=A_0$) and the $n=1$ mode. This approximation, not followed in \cite{cmplx}, captures the physical imprint of the log oscillations in several physical observables \cite{NewRev}, but here we will retain the full structure \Eq{logos}. The logarithmic oscillations are blurred out when we coarse grain the measure \Eq{qF} to scales $\gg\ell_\infty$. This coarse graining amounts to defining $y:=\ln|x/\ell_\infty|$ and taking the average of any function $f(y)$ \cite{NLM,frc2}:
\be\label{aver}
\langle f(y)\rangle:=\frac{1}{2\pi}\int_0^{2\pi}\rmd y\,f(y)\,,
\ee
which yields the constants
\be\label{mflr}
\langle F_\om\rangle=A_0\,,\qquad \langle F_\om^2\rangle=A_0^2+\sum_{n> 0}\frac{A_n^2+B_n^2}{2}\,.
\ee
Thus, if we drop the zero mode and set $A_0=0$, the profile
\be
\tilde F_\om(x):=\sum_{n>0}F_n(x)
\ee
reproduces Nottale's fractal lengths \Eq{notL} upon the identification 
\be
\zeta=\tilde F_\om\,.
\ee
Correspondingly, the relations \Eq{notlr} agree with \Eq{mflr}. If we insisted in having a negative average in the time direction as in \Eq{mflr} (a feature which we do not see as necessary, for the moment), then we would have to put direction labels $\mu$ on the amplitudes $A_n\to A_n^{(\mu)}$ and $B_n\to B_n^{(\mu)}$ and impose that both $A_n^{(0)}$ and $B_n^{(0)}$ are purely imaginary for all $n>0$ (or that, say, $A_n^{(0)}$ is purely imaginary and such that $|A_n^{(0)}|^2>B_n^{(0)2}$; these choices would modify the reality conditions in \Eq{cb} along the time direction). This would imply a complex-valued spacetime measure but no physical issue, provided all observables were computed by taking the average \Eq{aver}.

The last piece of the puzzle is the differentiability of \Eq{qF}. In general, $q(x)$ is differentiable everywhere except at a discrete infinity of points. However, special choices of the $n$-dependence of $A_n$ and $B_n$ can render $q(x)$ nowhere differentiable. To see this, we make yet another parametrization of the measure, inspired by the typical $n$-behavior of the coefficients $\xi_n$ and $\eta_n=\xi_n^*$ of Eq.\ \Eq{fen} found in complex and critical systems \cite{GlSo}:
\be\label{xin}
\xi_n=\xi\frac{\rme^{-\g n}}{n^u}\rme^{\rmi\psi_n}\,,
\ee
where $\xi$ is real and $n$-independent, $\g,u\geq 0$ parametrize an exponential or power-law behavior, and $\psi_n$ is a real $n$-dependent phase. Writing also $c_n=\exp(\rmi\b_n)$, where $\b_n:=n\om\ln(\ell_\infty/\ell_*)$, and comparing with Eq.\ \Eq{cb}, we get
\be\label{AB}
A_n=2\xi\frac{\rme^{-\g n}}{n^u}\cos(\psi_n+\b_n)\,,\qquad B_n=-2\xi\frac{\rme^{-\g n}}{n^u}\sin(\psi_n+\b_n)\,.
\ee
This expression allows us to make a prescription on the amplitudes $A_n$ and $B_n$ such that the measure \Eq{qF} is nowhere differentiable. In fact, in \cite{GlSo} it was found that functions of the form $g(x)=\sum_{n=0}^{+\infty} \xi_n x^{-s_n}$ are nowhere differentiable if the phases $\psi_n$ are random (more precisely, ergodic and mixing), which is the case provided $\psi_n$ varies fast enough with $n$. For instance, the phases $\psi_n=\Om$, $\psi_n=\Om n$ and $\psi_n=\Om\ln(\Om n)$, where $\Om$ is a constant, are too ``slow'' in $n$ and can produce at most a discrete infinity of singular points, while
\be\label{psin}
\psi_n=\Om n\ln(\Om n)\,,\qquad \psi_n = \Om n^2\,,\qquad \psi_n=\Om\rme^{n/\Om}\,,
\ee
or the solution of the recursive equation $\psi_{n+1}=\psi_n+a n$, where $a$ is irrational, all give rise to Weierstrass-type functions, which are nowhere differentiable \cite{GlSo}. Choosing the measure amplitudes in this way, we obtain the desired result: a stochastic (nowhere-differentiable) spacetime geometry with an intrinsic distance-time uncertainty, for any multi-fractional theory with measure \Eq{qF}.


\section{Discussion}\label{disc}


\subsection{Summary}

To summarize, a stochastic spacetime can be realized in multi-fractional theories by two inequivalent mechanisms:
\begin{itemize}
\item \emph{Multi-fractional derivatives} (section \ref{sec31}). In this case, realizing the stochastic integrals argument of \cite{trtls}, $T_\g$ enjoys the stochastic view independently of the choice of the measure $q(x)$ (with or without zero mode, with regular or random amplitudes). Nottale's scale relativity is not directly related to $T_\g$, but it could be a relative of $T_q$, which is an approximation of $T_\g$. 
\item \emph{Random measure} (section \ref{sec32}). All multi-fractional theories $T_v$, $T_q$ and $T_\g$ enjoy the stochastic view because they all share the same nowhere-differentiable measure $q(x)$ with random amplitudes. Nottale's scale relativity corresponds to the case where log oscillations average to zero (modulation function $\tilde F_\om$ in the measure).
\end{itemize}
Which option is better justified remains to be decided. The first one is valid only for the theory $T_\g$ with multi-fractional derivatives and does not require any \emph{Ansatz} for the measure amplitudes. The second one is valid for all multi-fractional theories, but it requires the \emph{Ansatz} \Eq{AB} with a fast varying $\psi_n$ such as the examples collected in \Eq{psin}. On the positive side, the choice \Eq{AB}, stemming from \Eq{xin}, was empirically found to be very general in complex and critical systems \cite{Sor98,GlSo}. But, to be fair, what holds in those branches of physics may not be valid in quantum gravity, where the nowhere differentiable function under scrutiny has a totally different role with respect to its counterparts in complex and critical systems. We do not know how to obtain \Eq{xin} and \Eq{psin} from first principles or from observations in multi-fractional theories, although cosmological constraints on $A_n$ and $B_n$ are under study \cite{cmplx}. Also, the mechanism we detailed for generating stochastic fluctuations of spacetime measurements might have consequences for the cosmological constant problem \cite{deC}.\footnote{No underlying theory was assumed in \cite{deC}, but the multi-fractional framework might provide theoretical justification of those results \emph{a posteriori}.}

Either way, if spacetime is stochastic, then the same measurement uncertainties calculated via heuristic arguments combining quantum mechanics and general relativity arise in multi-fractional theories. In this precise sense, classical multi-fractional theories encode quantum-gravity effects. 


\subsection{Related proposals}\label{relpro}

A relation between spacetime fuzziness and a fractal structure was suspected long since and it has been investigated under different perspectives during the years. By itself, this connection is not technically difficult to establish. For instance \cite{padma1}, it is sufficient to consider a metric formulation and deform the metric with corrections that depend on the geodesic distance $\s(x,x')$ between two points, but such that its zero-point length is non-vanishing, $\lim_{x'\to x}\s(x,x')\propto\lp$ \cite{Pa85a,Pa85b,Pad98}. Hence the four-volume is deformed. In this set-up, which is similar to the one in rainbow gravity, one can choose the corrections so that to obtain simultaneously a varying Hausdorff dimension and a spacetime uncertainty \cite{padma1}. The real challenge, however, is to embed such connection in a top-down theory and to explain its physical origin.

The first datum we would like to mention does not link fuzziness and multi-scale spacetimes explicitly but it provides indirect support of the above view that a classical stochastic spacetime efficiently reproduces quantum-gravity effects. Coordinates defined on a nowhere-differentiable geometry obey an uncertainty principle virtually identical to Heisenberg's \cite{BAP}. In other words, the nowhere-differentiable structure typical of multi-fractional theories (with multi-fractional derivatives or a random measure) naturally reproduce quantum-mechanical effects such as those considered in section \ref{revi}. If we recall that nowhere differentiability is typical of sets with non-integer dimension \cite{frc1,KoG,RYS,RLWQ,NLM}, then the relation between measurement uncertainty and dimensional flow becomes apparent.

On the other hand, intrinsic spacetime fuzziness is the starting point of non-commuta\-tive spacetimes \cite{DFR1}. There, the idea is to get QG$=$QM$+$GR from a fuzzy spacetime rather than the latter from the former. Getting quantum-gravity as a byproduct of a non-trivial integro-differential structure is also the path followed of multi-fractional theories, which made us wonder about possible connections between non-commutative spacetimes and the multi-fractional paradigm \cite{ACOS,CR1}. Despite a number of similarities in dimensional flow, there is no quantitative connection between these two frameworks, mainly because in the former coordinates do not factorize in effective measures. However, in the present paper we have finally found the reason beyond those similarities: it is because dimensional flow and distance-time uncertainties have the same origin in both theories. In the case of fuzziness coming from a random measure (section \ref{sec32}), this common origin is the quasi-universality of dimensional flow, established by the first flow-equation theorem for non-factorizable geometries and by the second theorem for factorizable ones \cite{NewRev,first}.

Fuzziness understood as a spacetime foam was related to a multi-scale quan\-tum-gravity structure already in \cite{CrSm1,CrSm2}. As a model of spacetime ``foam,'' Crane and Smolin took a scale-invariant distribution of Planckian black holes. If one then considers a perturbative quantization of gravity, this is sufficient to deform the dimension dependence of the graviton propagator and to improve renormalizability of the theory. Two differences with respect to our approach are the implementation of general-relativistic features by hand (in this case, microscopic black holes) and the derivation from there of an anomalous (or multi-scale, or fractal) spacetime structure. Quantum mechanics and general relativity were joined there (in the form of a black-hole foam) \emph{ad hoc}, which is essentially the same philosophy of the estimates reviewed in section \ref{revi}. However, a notable upgrade by \cite{CrSm1,CrSm2} with respect to arguments of \cite{ngdam,amelino} is the recognition that the resulting spacetime uncertainty is responsible for introducing a scale hierarchy, making spacetime multi-scale or fractal. Here we proceeded the other way around, taking a spacetime which is multi-scale by default and getting a stochastic structure (in turn determined by the integral or differential calculus realizing multi-scaling) from that, using a very minimal list of ingredients: unique parametrization of the spacetime measure from the flow-equation theorem and randomization of the measure amplitudes as suggested by results from critical and complex systems.

In \cite{Co2}, a phenomenological dispersion relation was proposed to recover the running profile of the spectral dimension found numerically in causal dynamical triangulations. The same dispersion relation was then employed to write down the expression of the geodesic distance between two points, which happens to depend on the resolution of the probe. Thus, in causal dynamical triangulations one can get fuzziness from dimensional flow. In that paper, this connection is, to our understanding, not completely explicit and there is no direct reference to fuzziness, but there is a more pressing issue one should be careful about. Inferring a modified dispersion relation from a given dimensional flow is a risky procedure that likely incurs in the twin problem \cite{CES} well known in transport theory \cite{Sok12}, stating that very different diffusion equations can give rise to the same asymptotic form of the return probability. In quantum gravity, this means that one can get the same dimensional flow (up to irrelevant differences in transient regimes) from very different diffusion processes, Laplacians, and diffusion operators \cite{frc4}. Thus, the result of \cite{Co2} relies on an intermediate step (the guess of a deformed dispersion relation) whose physical grounds are not clear to us. Nevertheless, it may provide circumstantial evidence of the relation between dimensional flow and fuzziness in causal dynamical triangulations.

Of all these examples, multi-fractional theories, non-commutative spacetimes and (with the above reservations) causal dynamical triangulations are top-down examples; the others rely on isolated theoretical observations or on the heuristics of quantum gravity.


\subsection{Avoiding observational constraints on multi-fractional theories}

We feel confident that the observations we made here might represent an important step towards understanding why the running of dimensions at short scales is a universal property of QG approaches. In particular, we have argued that dimensional flow is linked to distance-time fuzziness, whose form can be inferred from arguments combining quantum mechanics and general relativity, without knowledge of the detailed features of one or another QG model. In this way, we have been able to pick out two preferred values for the fractional exponent of the measure $\alpha$. If we take seriously the parameter fixing suggested by the QM$+$GR arguments and their correspondence with multi-scale spacetimes, then we can refine previous bounds on $\ell_*$, $t_*$ and the associated energy scale $E_*$. The $\a=1/2$ case has already been considered in the literature, while the other is reported in table \ref{tab1} for the most effective experiments or observations by which the theory with $q$-derivatives has been tested.\footnote{We do not have bounds on $T_\g$ yet.}
\begin{table}[ht!]
\begin{center}
\caption{\label{tab1}Bounds on the hierarchy of the multi-fractional theory $T_q$ with $q$-derivatives for $\a_0=1/3=\a$ and $\a_0=1/2=\a$. Bounds from the Lamb shift and from gravitational waves refer to the most conservative estimates with generic coefficients in the correction terms (see \cite{NewRev}, especially table 8, for details). ``Pseudo'' indicates bounds obtainable only in the stochastic view (which is an approximated step in $T_q$) and only in the case where photon-graviton propagation speeds differ by a maximal random fluctuation. There is no useful bound on the measurement and variation of the fine-structure constant $\a_\textsc{qed}$ \cite{frc13}. For $\a=1/2$, there are also bounds on the amplitudes $A_1$ and $B_1$ in \Eq{logosb} \cite{frc14}. Bounds without references have been obtained in this paper.}
\begin{tabular}{|l|ccc|c|c|}\hline
$T_q$	($\a_0=1/3=\a$)	& $t_*$ (s)     		& $\ell_*$ (m) & $E_*$ (GeV)& Avoided in  \\
											&										&							 &						& stochastic view\\\hline\hline
CMB black-body				& $<10^{-23}$       & $<10^{-15}$  & $>10^{-1}$ & only for $A_0=0$\\
spectrum 							&										&							 &						& only for $A_0=0$\\
Lamb shift        		& $<10^{-24}$       & $<10^{-15}$  & $>10^{-1}$ & only for $A_0=0$\\									
Gravitational waves		& $<10^{-32}$       & $<10^{-24}$  & $>10^{7}$  & also for $A_0\neq0$\\
(pseudo)							&										&							 &						& \\
GRBs 				 					& $<10^{-47}$       & $<10^{-39}$  & $>10^{23}$ & only for $A_0=0$\\
Vacuum Cherenkov   	  & $<10^{-68}$     	& $<10^{-60}$  & $>10^{44}$ & only for $A_0=0$\\
radiation							&										&							 &						& \\\hline
$T_q$	($\a_0=1/2=\a$)	& $t_*$ (s)     		& $\ell_*$ (m) & $E_*$ (GeV)& Avoided in  \\
											&										&							 &						& stochastic view\\\hline\hline
CMB black-body      	& $<10^{-26}$       & $<10^{-18}$  & $>10$      & only for $A_0=0$ \\
spectrum \cite{frc14}	&										&							 &						& only for $A_0=0$\\
Lamb shift \cite{frc13,NewRev}& $<10^{-26}$       & $<10^{-18}$  & $>10$ & only for $A_0=0$\\									
Gravitational waves		& $<10^{-42}$       & $<10^{-33}$  & $>10^{17}$ & also for $A_0\neq0$\\
(pseudo) \cite{qGW,NewRev} &										&							 &						& \\
GRBs \cite{qGW}				& $<10^{-57}$       & $<10^{-48}$  & $>10^{32}$ & only for $A_0=0$\\
Vacuum Cherenkov 			& $<10^{-79}$     	& $<10^{-71}$  & $>10^{55}$& only for $A_0=0$\\
radiation	\cite{NewRev}						&										&							 &						& \\\hline
\end{tabular}
\end{center}
\end{table}

While bounds from the cosmic microwave background (CMB) black-body spectrum and from the Lamb shift in quantum electrodynamics change only by two orders of magnitude from one case to the other, the observation of black-hole gravitational waves is more sensitive to the value of the fractional exponents. For $\a=1/2=\a_0$, the energy $E_*$ is not much smaller than the grand-unification scale \cite{qGW}, while for $\a=1/3=\a_0$ it is $E_*>10^4\,\text{TeV}$, 1000 times larger than the LHC run-2 center-of-mass energy. The bounds from gamma-ray bursts (GRB) have been determined much less rigorously \cite{qGW}. As already known, they exclude the $\a=1/2=\a_0$ case because $E_*>10^{13}\mpl$. For $\a=1/3=\a_0$, this bound is less severe but still above the Planck mass, $E_*>10^4\mpl$. A detailed calculation of the effect of multi-fractional geometries in $T_q$ and $T_\g$ on the propagation of high-energy photons in a cosmological background will be needed to check whether these estimates are robust, although there seems to be little hope at least for $T_q$ \cite{qGW}. Since $T_q$ can be regarded also as an approximation of $T_\g$ \cite{NewRev}, observational bounds on $T_\g$ could be conjectured to be very similar to those on $T_q$. We know much less about $T_\g$ and we cannot rule it out with our present theoretical understanding of it. Four things may happen that could save the theory $T_\g$: (i) that the GRB and vacuum Cherenkov radiation bounds are somehow flawed under a closer scrutiny; (ii) that, despite their similarities, $T_q$ and $T_\g$ are essentially different in some key physical consequences, as also technical reasons seem to indicate \cite{NewRev}; (iii) that the fractional derivatives in $T_\g$ must or can be taken with an order $\g$ smaller than the fractional exponent $\a$ in the measure; (iv) that the heuristic arguments of \cite{ngdam,amelino} do not fix the fractional exponents as claimed in this paper; or (v) that these arguments do fix the fractional exponents and the stochastic view holds with no zero mode in the measure ($A_0=0$). Case (v) would also save $T_q$. The most likely possibilities are, in our opinion, (ii) and (v). Case (ii) is the most attractive to us, but only explicit calculations will be able to check it. Regarding (v), in the stochastic view the averaging to zero of stochastic fluctuations in the propagation of particles can easily avoid all constraints (including from gravitational waves, GRBs and vacuum Cherenkov radiation) coming from modified dispersion relations, which are the strongest to date. However, in this case it would be difficult to falsify $T_q$ and $T_\g$. Letting $A_0\neq 0$ would avoid the gravitational-wave bound but not those from GRBs and Cherenkov radiation. Case (iv) may still be possible and one would consider it only in the deterministic view, which would amount to dissociate the heuristic quantum-gravity arguments from multi-fractional theories.


\subsection{Structure of multi-scale spacetimes}

We make some remarks on the structure of spacetime uncovered here for the theory $T_\g$ and its approximation $T_q$. The pair of equations \Eq{l13}--\Eq{t13} on one hand and \Eq{l12}--\Eq{t12} on the other hand describe two different geometries. In this concluding section, we comment on both. The only feature in common is scaling isotropy, for which the spectral and Hausdorff dimensions follow the same UV running, as noted in the introduction. It is intriguing that, as a byproduct of our analysis, we found such a property. 
\begin{itemize}
\item $\a=\tfrac13=\a_0$. A characteristic worth special mention for this case is the identification of the scales $t_*$ and $\ell_*$ with the Planck time $\tp$ and length $\lp$. The main consequence of this finding is that the binomial measure with log oscillations is not just an approximation of a more complicated multi-scale polynomial measure: since there is no meaning to scales below $\ell_*=\lp$, there is no other scale than $\ell_*$ in the hierarchy of the theory. A mild theoretical support to these features is what we might call a ``scale equipartition.'' In deriving Eq.\ \Eq{qF}, we introduced the arbitrary scale ratio \Eq{ceqi}, corresponding to a non-zero constant phase $\b_n$ in the amplitudes \Eq{AB}. However, one might as well impose equipartition of the same fundamental length in power-law and oscillatory terms:
\be
\ell_\infty=\ell_*\qquad \Longrightarrow \qquad c_n=1\,,\quad \b_n=0\,.
\ee
Taking on board results in non-commutative spacetimes that indicate that $\ell_\infty=\lp$ \cite{ACOS,CR1}, we get
\be\label{lll}
\ell_\infty=\ell_*=\lp\,,
\ee
in agreement with Eq.\ \Eq{l13}. The case of the time direction is similar. Note that this spacetime is not normed in the UV of the theory $T_\g$, since $\a<1/2$ \cite{NewRev,frc1}. This is not a problem in the stochastic view, where intervals lose meaning anyway at scales $\sim \ell_*$. The transition from a normed deterministic space to a fuzzy one is gradual and nothing special happens exactly at the scale $\ell_*$. To summarize, we end up with a binomial isotropic-scaling geometry with $\a=1/3=\a_0$ and one fundamental \emph{absolute} scale \Eq{lll}, which marks a smooth transition from a normed spacetime to a non-normed fuzzy spacetime. Interestingly, having $\a_0=1$ (ordinary time direction) but $\a=1/3$ corresponds to a spectral dimension $\ds\simeq 1+3/3=2$ in the UV, the same asymptotic configuration of Ho\v{r}ava--Lifshitz gravity \cite{Hor3}. This is not a coincidence, since the critical scaling of coordinates in Ho\v{r}ava--Lifshitz gravity is easily reproduced by geometric coordinates \cite{NewRev,fra7}.
\item $\a=\tfrac12=\a_0$. This is nothing but the special value $1/2$ of the multi-fractional literature, the minimum exponent at which a norm exists in $T_\g$ \cite{NewRev,fra1,frc1}. The main point of departure from the $\a=1/3=\a_0$ case is that here a norm does exist in the UV, but it is not unique. This is the so-called Manhattan or taxicab geometry, where two points can be connected by many different geodesic paths with the same minimum length \cite{frc1}. Therefore, in this case spacetime is normed at all scales but geodesics lose uniqueness in the deep UV.

On top of that, in this second case $\ell_*$ (and $t_*$) shows a dependence on the scale $s$ (or $t$) at which spacetime events are probed. At macroscopic scales $s \gg \lp$ ($t \gg \tp$), $\ell_* \rightarrow 0$ ($t_* \rightarrow 0$) and there is no dimensional flow ($\de L_\a\to 0$). On the other hand, lowering $s$ ($t$), the scale $\ell_*$ ($t_*$) increases up to $\lp$ ($\tp$) when we are measuring geometry or testing spacetime events at Planckian distances. This dependence of $\ell_*$ (or $t_*$) on the observation scale reveals a crucial and often advertized feature of multi-fractional theories, namely, the fact that in a multi-scale geometry measurements depend on the scale at which the experiment is being performed. Thus, Eq.\ \eqref{l12} agrees with the perspective according to which multi-fractional models are theories where the result of measurements is affected by the scale $s$ ($t$) of the observer. In this sense, we can talk about a \emph{relative} multi-scale hierarchy among observers, although this does not exclude the existence of an absolute hierarchy as in the previous case. Imposing the phenomenological limitations on measurements of section \ref{revi} to hold only at the Planck scale ($s=\lp$, $t=\tp$), we recover an absolute hierarchy and the identification \Eq{lll}.
\end{itemize}
All this combines with the microscopic discrete scale invariance of the measure \Eq{logos} \cite{NewRev,frc2} to give non-smooth geometries that can describe the deepest UV recesses of quantum gravity and that will deserve further study. 


\subsection{Conclusions}

We established a relation between dimensional flow and spacetime fuzziness by working in the framework of multi-fractional theories. The main reason why we focused on these models is that they are easy to manipulate. However, this correspondence does not hold exclusively for multi-fractional geometries, as argued in \cite{ACCR}. The approach of multi-fractional spacetimes just provides a first exploratory study useful to recognize a striking feature that may be much more general and characteristic of other QG theories. We are aware that testing our conjecture may be harder in other formalisms of quantum gravity, but we feel confident that the encouraging results reported here will energize efforts in that direction. All the main elements of our arguments are already in place in some of the major proposals of the literature. We commented on non-commutative spacetimes and causal dynamical triangulations in section \ref{relpro}, but there is more. In particular, both asymptotically-safe quantum gravity and the discrete-geometry, mutually related frameworks of loop quantum gravity, spin foams and group field theory have dimensional flow \cite{NiR,RSnax,COT2,COT3,LaR5,dimLQG} and implement fuzziness by the presence of minimal lengths, areas or resolutions \cite{rov07,RSc1,RSc2}. However, although a relation between anomalous dimension and fuzzy features certainly seems to exist in these cases, so far it has been at best indirect or purely technical. Revisiting these theories in search of a physical connection similar to that found here may help to clarify some of their formal aspects and even give new tools by which to extract testable phenomenology.


\section*{Acknowledgments}
G.C.\ is under a Ram\'on y Cajal contract and is supported by the I+D grant FIS2014-54800-C2-2-P. M.R.\ thanks Instituto de Estructura de la Materia (CSIC) for the hospitality during an early elaboration of this work. This article is based upon work from COST Action CA15117, supported by COST (European Cooperation in Science and Technology).




\begin{thebibliography}{99}
\bibitem{Smo17} L.\ Smolin, \tia{What are we missing in our search for quantum gravity?} \arX{1705.09208}.
\end{thebibliography}
\end{document}